\documentclass[epj]{webofc}
\usepackage[varg]{txfonts}   % Web of Conferences font
\wocname{EPJ Web of Conferences}
\woctitle{CONF12}
\newcommand{\be}{\begin{eqnarray}}
\newcommand{\ee}{\end{eqnarray}}
\def\ra{\rangle}
\def\la{\langle}

  \newcommand{\exclude}[1]{}

\newcommand{\beq}{\begin{equation}}
\newcommand{\eeq}{\end{equation}}

\begin{document}
\selectlanguage{english}
\title{Beyond WIMPs: the Quark (Anti) Nugget Dark Matter}
%
% subtitle (optional, strongly discouraged)

\author{Ariel Zhitnitsky\inst{1}\fnsep\thanks{\email{arz@phas.ubc.ca}}  }

\institute{Department of Physics and Astronomy, University of
  British Columbia, Vancouver,  Canada
 }

\abstract{%
We review   a   testable 
dark matter (DM) model outside of the standard WIMP paradigm. The model is unique in a sense that  the 
observed ratio $\Omega_{\rm dark} \simeq   \Omega_{\rm visible}$ for 
visible and dark matter 
densities  finds its natural explanation  as a result
of their common QCD origin when both types of matter (DM and visible) are formed during  the QCD transition and both are proportional to single dimensional parameter of the system, $\Lambda_{\rm QCD}$. We argue that the  charge separation effect also  inevitably  occurs during the same QCD transition in the presence of the  $\cal{CP}$ odd axion field $a(x)$. It  leads  
to preferential formation of one species of nuggets on the  scales of the visible Universe where the axion field $a(x)$ is coherent. A natural outcome of  this preferential evolution   is that only one type of the visible baryons (not anti- baryons) remain in the system after the nuggets complete their formation.   Unlike conventional WIMP dark matter candidates, the   nuggets and anti-nuggets are strongly interacting but macroscopically large objects. The rare events of annihilation of the anti-nuggets with visible matter lead to a number of observable effects. 
We argue that the relative intensities for a number of measured 
excesses of emission  from the centre of galaxy (covering more than 11 orders of magnitude)   are determined by  standard and well established physics. At the same time    the  absolute intensity of emission is determined by a single new fundamental parameter of the theory, the axion mass,    $10^{-6} {\rm eV} \lesssim  m_a \lesssim  10^{-3}{\rm eV}$.  Finally, we  comment on implications of  these studies  for   the  axion search experiments, including microwave cavity  and the Orpheus experiments. 

}
\maketitle
\section{Introduction}
\label{intro}
This talk is mostly based on recent paper \cite{Liang:2016tqc}.
It is generally assumed that the Universe 
began in a symmetric state with zero global baryonic charge 
and later, through some baryon number violating process, 
evolved into a state with a net positive baryon number. As an 
alternative to this scenario we advocate a model in which 
``baryogenesis'' is actually a charge separation process 
in which the global baryon number of the Universe remains 
zero. In this model the unobserved antibaryons come to comprise 
the dark matter in form of the dense heavy nuggets, similar to the Witten's strangelets \cite{Witten:1984rs}.
 Both quarks and antiquarks are 
thermally abundant in the primordial plasma but, in 
addition to forming conventional baryons, some fraction 
of them are bound into heavy nuggets of quark matter.   
  Nuggets of both matter and 
antimatter are formed as a result of the dynamics of the axion domain walls \cite{Zhitnitsky:2002qa,Oaknin:2003uv,Liang:2016tqc}, some details of this process will be discussed  
later in the text. 

An overall coherent baryon asymmetry in the entire Universe is  a result of the strong CP violation due to the fundamental $\theta$ parameter in QCD which is assumed to be nonzero at the beginning of the QCD    transition\footnote{It is known that that 
the QCD transition is actually a crossover rather than a phase transition \cite{Aoki:2006we}. 
In context of the present paper the important factor is the scale $\sim 170$ MeV where transition  happens rather than its precise nature.}.  This source of strong CP violation is no longer 
available at the present epoch as a result of the axion dynamics, see original  \cite{axion,KSVZ,DFSZ} and more
 recent papers \cite{vanBibber:2006rb, Sikivie:2008, Asztalos:2006kz,Raffelt:2006cw,Sikivie:2009fv,Rybka:2014cya,Rosenberg:2015kxa,Graham:2015ouw} on the subject.
Were CP symmetry to be exactly preserved  
an equal number of matter and antimatter nuggets would form resulting in 
no net ``baryogenesis". However, CP violating processes associated 
with the axion $\theta(x)$ term in QCD result in the preferential formation of 
one type of species. 
This preference  is essentially determined by the dynamics of  coherent axion field   ${\theta}(x)$ at the initial stage of the nugget's formation.   

Asymmetric production of the nuggets directly translates into asymmetry between visible baryons and anti-baryons
because the total baryon charge is a conserved quantity in this framework. If more anti-nuggets are produced in the system then  less 
conventional anti-baryons remain in the system.   
These remaining anti-baryons in the plasma then 
annihilate away leaving only the baryons whose antimatter 
counterparts are bound in the excess of anti-nuggets and thus 
unavailable to annihilate. One should emphasize that the resulting asymmetry is order of one.  It is not sensitive to a relatively small magnitude of the axion mass\footnote{The axion's mass $m_a (T)$ as a function of the temperature
 has been computed using  the lattice simulations \cite{Borsanyi:2016ksw} with very high accuracy in the region  well above the QCD transition.} nor  to the relatively small magnitude of the axion filed ${\theta}(x)\sim (10^{-2}-10^{-4})$ at the beginning of the QCD transition as long as it remains  coherent on the scale of the Universe, see \cite{Liang:2016tqc}  for the details.  
  This  is precisely  the main reason of why the  visible and dark matter 
densities must be the same order of magnitude 
\be
\label{Omega}
 \Omega_{\rm dark} \approx \Omega_{\rm visible}
\ee
   as they both  proportional to the same fundamental $\Lambda_{\rm QCD} $ scale,  and they both are originated at the same  QCD epoch.  In particular, if one assumes that the nuggets and anti-nuggets saturate the dark matter density today than 
  the observed 
matter to dark matter ratio $\Omega_{\rm dark} \simeq 5\cdot \Omega_{\rm visible}$ corresponds to a specific proportion  when  number  of anti-nuggets is larger than number of nuggets 
  by a factor of $\sim$ 3/2 at the end of nugget's formation. This would 
result in a matter content with baryons, quark nuggets 
and antiquark nuggets in an approximate  ratio 
\be
\label{ratio1}
|B_{\rm visible}|: |B_{\rm nuggets}|: |B_{\rm antinuggets}|\simeq 1:2:3, 
\ee
 with  no net baryonic charge. 
 If these processes 
are not fundamentally related the two components $\Omega_{\rm dark}$ and $\Omega_{\rm visible}$  could easily 
exist at vastly different scales.

Unlike conventional dark matter candidates, dark-matter/antimatter
nuggets are strongly interacting but macroscopically large.  
They do not contradict the many known observational
constraints on dark matter or
antimatter  for three main reasons~\cite{Zhitnitsky:2006vt}:
\begin{itemize} 
\item They carry a huge (anti)baryon charge 
$|B|  \gtrsim 10^{25}$, and so have an extremely tiny number
density; 
\item The nuggets have nuclear densities, so their effective interaction
is small $\sigma/M \sim 10^{-10}$ ~cm$^2$/g,  well below the typical astrophysical
and cosmological limits which are on the order of 
$\sigma/M<1$~cm$^2$/g;
\item They have a large binding energy 
such that the baryon charge  in the
nuggets is not available to participate in big bang nucleosynthesis
(\textsc{bbn}) at $T \approx 1$~MeV. 
\end{itemize} 
To reiterate: the weakness of the visible-dark matter interaction 
in this model due to the small geometrical parameter $\sigma/M \sim B^{-1/3}$ 
rather than due to the weak coupling 
of a new fundamental field to standard model particles. 
It is this small effective interaction $\sim \sigma/M \sim B^{-1/3}$ which replaces 
the conventional requirement of sufficiently weak interactions for WIMPs.

A fundamental measure of the scale of baryogenesis is the 
baryon to entropy ratio at the present time
\be
\label{eta}
\eta\equiv\frac{n_B-n_{\bar{B}}}{n_{\gamma}}\simeq \frac{n_B}{n_{\gamma}}\sim 10^{-10}.
\ee
If the nuggets were not present after the transition the conventional baryons 
and anti-baryons would continue to annihilate each other until the temperature 
reaches $T\simeq 22$ MeV when density would be 9 orders of magnitude smaller 
than observed. This annihilation catastrophe, normally thought to be  resolved as a result of  ``baryogenesis," is avoided in our proposal because  more  anti-baryon charges  than 
baryon charges  are hidden in the form of the macroscopical nuggets and thus no longer available 
for annihilation. Only the visible baryons (not anti-baryons)  remain in the system 
after nugget formation is fully completed.

In our proposal (in contrast with conventional models) the ratio $\eta$ is determined 
by the formation temperature $T_{\rm form}$ at 
which the nuggets and anti-nuggets basically 
have completed  their formation and below which annihilation with 
surrounding matter becomes negligible.    
This temperature is determined by many factors: transmission/reflection 
coefficients, evolution of the nuggets, expansion of the universe, cooling rates, evaporation 
rates, the  dynamics of the axion domain wall network, etc. 
In general, all of these effects will contribute   equally to 
determining $T_{\rm form}$ at  the QCD scale. Technically, the corresponding 
effects are hard    to compute as even basic properties of the  QCD phase diagram at nonzero 
$\theta$ are still unknown. 
\begin{figure}[h]
\centering
\sidecaption
\includegraphics[width= 4cm,clip]{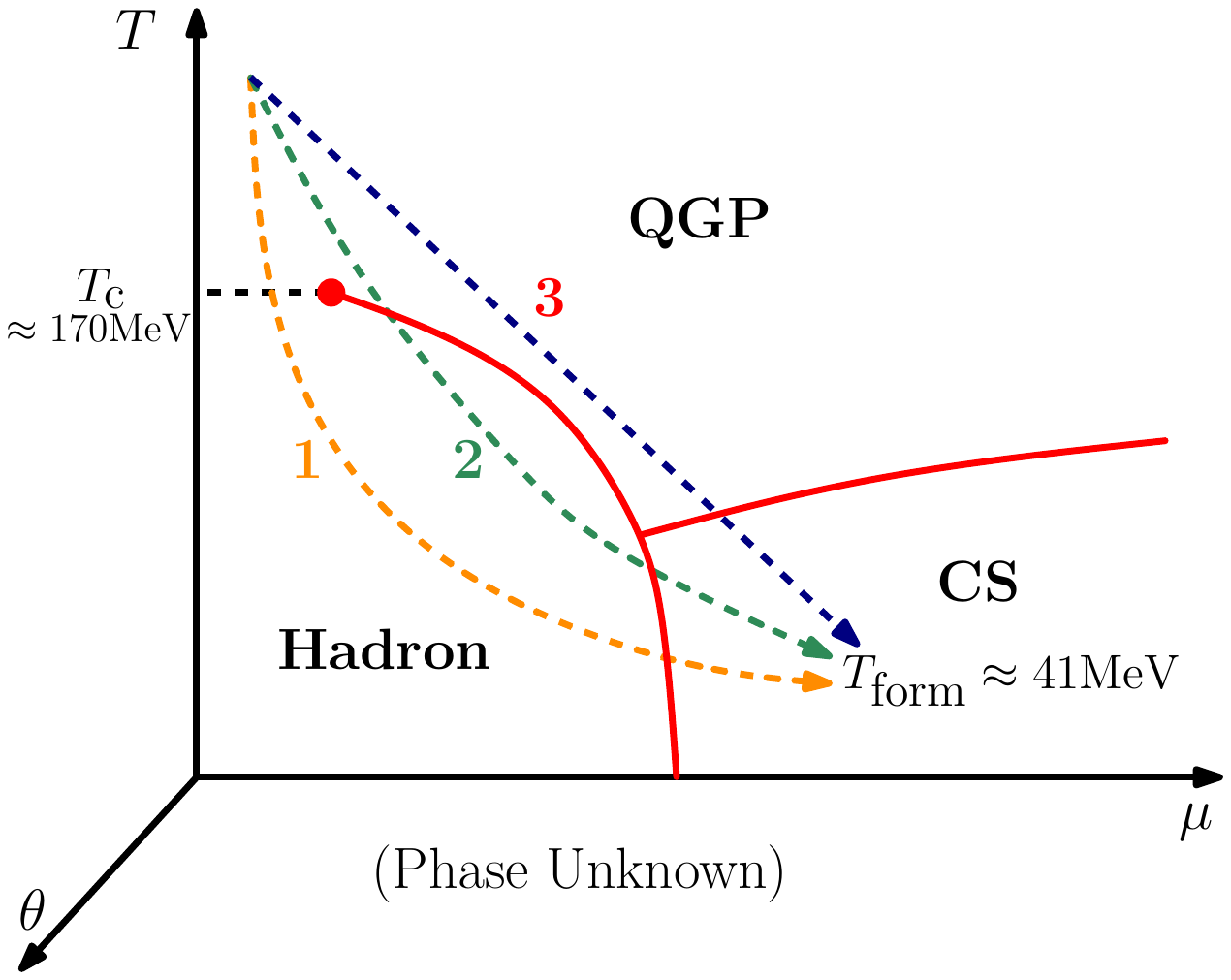}
\caption{ The conjectured phase diagram.  The plot is taken from \cite{Liang:2016tqc}. Possible cooling paths are denoted as path 1, 2 or  3. The phase diagram is in fact much more complicated as the dependence on the third essential parameter, the $\theta$ is not shown as it is largely unknown.   It is assumed that the final destination after the nuggets are formed is the CS region with $T_{\rm form}\approx 41$ MeV, $\mu >\mu_c$ and $\theta\approx 0$, corresponding to the presently observed ratio (\ref{eta}), see text for the details.  }
\label{fig:phase_diagram}
\end{figure}
However, an approximate estimate of $T_{\rm form}$  is 
quite simple as it must be  expressed in terms of the gap 
$\Delta\sim 100$ MeV when the colour 
superconducting  phase  sets in  inside the nuggets. The observed ratio (\ref{eta}) 
corresponds to $T_{\rm form}\simeq 41$ MeV 
which is  indeed a typical QCD scale slightly below the critical 
temperature $T_c\simeq 0.6 \Delta$ when colour superconductivity (CS) sets in. 
In different words, in this proposal the ratio (\ref{eta}) emerges as a result of the 
QCD dynamics when process of charge separation stops at  $T_{\rm form}\simeq 41$ 
MeV, rather than a result of baryogenesis when a net baryonic charge is produced.

\section{ Quark (anti) nugget DM    confronting the observations}\label{nuggets}

 While the observable consequences of this model are on average strongly suppressed  
by the low number density  of the quark nuggets $\sim B^{-1/3}$ as explained above, the interaction of these objects 
with the visible matter of the galaxy will necessarily produce observable 
effects. Any such consequences will be largest where the densities 
of both visible and dark matter are largest such as in the 
core of the galaxy or the early universe. In other words, the nuggets behave as  a conventional cold DM in the environment where density of the visible matter is small, while they become interacting and emitting radiation  objects (i.e. effectively become visible matter)  when they are placed in the environment with sufficiently large density.

 The relevant phenomenological features of the resulting nuggets 
 are determined by properties of the so-called electro-sphere  as discussed in original refs. \cite{Oaknin:2004mn, Zhitnitsky:2006tu,Forbes:2006ba, Lawson:2007kp,Forbes:2008uf,Forbes:2009wg}. 
 These properties are in principle, calculable from first principles using only 
the well established and known properties of QCD and QED. As such 
the model contains no tunable fundamental parameters, except for a single mean baryon number  $\la B\ra$
which itself is determined by the axion mass $m_a$  as we already mentioned.

 A comparison between   emissions with drastically different frequencies 
from the centre  of galaxy is possible because the rate of annihilation events (between visible matter and antimatter DM nuggets) is proportional to 
 the product of the local visible and DM distributions at the annihilation site. 
The observed fluxes for different emissions thus depend through one and the same line-of-sight integral 
\be
\label{flux1}
\Phi \sim R^2\int d\Omega dl [n_{\rm visible}(l)\cdot n_{DM}(l)],
\ee
where $R\sim B^{1/3}$ is a typical size of the nugget which determines the effective cross section of interaction between DM and visible matter. As $n_{DM}\sim B^{-1}$ the effective interaction is strongly suppressed $\sim B^{-1/3}$ as we already mentioned  in the Introduction. The parameter $\la B\ra\sim 10^{25}$  was fixed in this  proposal by assuming that this mechanism  saturates the observed  511 keV line   \cite{Oaknin:2004mn, Zhitnitsky:2006tu}, which resulted from annihilation of the electrons from visible matter and positrons from anti-nuggets. It has been also assumed that the observed dark matter density is saturated by the nuggets and anti-nuggets.  It corresponds to an average baryon charge  $\la B\ra \sim 10^{25}$ for typical  density distributions  $n_{\rm visible}(r),  n_{DM}(r)$ entering (\ref{flux1}). Other emissions from different bands  are expressed in terms of the same integral (\ref{flux1}), and therefore, the  relative\textsc{} intensities  are completely determined by internal structure of the nuggets which is described by conventional nuclear physics and basic QED. We present a short overview of these results below.  

Some  galactic electrons are able to penetrate to a sufficiently 
large depth of the anti-nuggets. These events
 no longer produce  the characteristic positronium decay 
spectrum (511 keV line  with a typical width of order $\sim {\rm few~ keV}$ 
accompanied by the conventional continuum due to $3\gamma$ decay)   but a direct non-resonance $e^-e^+ \rightarrow 2\gamma$ emission spectrum. 
 The transition between the resonance positronium decays and non-resonance 
  regime is determined by conventional physics and allows us to  compute   
the strength and spectrum of the MeV scale emissions relative to 
that of the 511~keV line \cite{Lawson:2007kp,Forbes:2009wg}.  Observations 
by the \textsc{Comptel} satellite indeed show some  excess above the galactic 
background  consistent with our estimates. 

Galactic protons incident on the anti-nugget 
will penetrate some distance into the quark matter before 
annihilating into hadronic jets. This process results in the emission of Bremsstrahlung 
photons at x-ray energies \cite{Forbes:2006ba}. Observations by the 
\textsc{Chandra} observatory apparently  indicate an excess in x-ray emissions from 
the galactic centre.  
Hadronic jets produced 
deeper in the nugget or emitted in the downward direction 
will be completely absorbed. They  eventually emit 
thermal photons with radio frequencies \cite{Forbes:2008uf,Lawson:2012zu,Lawson:2015xsq}. Again the relative scales of these 
 emissions may be estimated and is found to be in 
agreement with observations.  

   We conclude this brief overview on  observational constraints of the model with the following remark.  This model which has  a single  fundamental parameter  (the  mean baryon number of a nugget $\la B\ra \sim 10^{25}$, corresponding to the axion mass $m_a\simeq 10^{-4}$ eV), and which      enters  all the computations is   consistent with all known astrophysical, cosmological, satellite  and ground based constraints as highlighted above. Furthermore, in a number of cases the predictions of the model are very close to the presently available  limits, and very modest improving of those constraints may lead to a discovery of the nuggets.  Even more than that: there is a number of frequency bands where some excess of emission was observed, and this model may explain some portion, or even entire excess of the observed radiation in these frequency bands.
 \exclude{
 In the light of  this (quite optimistic) 
 assessment  of the  observational constraints   of this model  it is  quite obvious    that   further and deeper studies  of this model    are worthwhile to pursue. The relevant developments may   include, but not limited, to such hard problems   as formation mechanisms during the QCD transition in early Universe,  even though many  key elements for proper addressing those questions at $\theta\neq 0, \mu\neq 0, T\neq 0$ are still  largely unknown in strongly coupled QCD as shown on Fig.\ref{fig:phase_diagram}. This work is the first step in the direction to explore a possible mechanism of formation of the nuggets.  
       }
\section{Five   crucial  ingredients of the proposal.}\label{formation}
In this section we explain  the  crucial elements of the proposal. The detail discussions for each ingredient can be found in original paper  \cite{Liang:2016tqc}.
\subsection{$N_{DW}=1$ domain walls}\label{DW}
  First important element of this proposal  is the presence of the topological objects, the axion domain walls \cite{Sikivie}.  As we already mentioned the  $\theta$ parameter is   the angular variable, and therefore supports various types of the domain walls, including the so-called $ N_{DW}=1$ domain walls when $\theta$ interpolates between one and the same physical vacuum state with the same energy $\theta\rightarrow\theta+2\pi n $.    The axion domain walls may form at the same moment when the axion potential get tilted, i.e. at  the moment $T_a$ when the axion field starts to roll due to the misalignment mechanism. The tilt becomes much more pronounced  at the transition when the chiral condensate forms at $T_c$.
In general one should expect that the $ N_{DW}=1$ domain walls form once the axion potential is sufficiently tilted, i.e. anywhere between $T_a$ and $T_c$.

One should comment here that it is normally assumed that for the topological defects to be formed the Peccei-Quinn (PQ)  phase transition must occur after inflation. This argument is valid for a generic type of  domain walls with  $ N_{DW}\neq 1$.  The conventional argument is    based on  the fact that  few physically  {\it different vacua} with the same energy must be present inside of the same horizon for the domain walls to be formed. The $ N_{DW}=1$ domain walls
are unique and very special in the  sense that $\theta$ interpolates between  {\it one and the same} physical vacuum state.   Such $ N_{DW}=1$ domain walls can be formed even if the PQ phase transition occurred before inflation and a unique physical vacuum occupies entire Universe \cite{Liang:2016tqc}.

 It has been  realized many years after  the original publication   \cite{Sikivie}       that  the  axion domain walls, in general, demonstrate a  sandwich-like  substructure
on  the QCD scale $\Lambda_{QCD}^{-1}\simeq $ fm.
The arguments supporting the 
QCD scale substructure inside the axion domain walls are based on analysis \cite{FZ}
of QCD in the large $N$ limit with inclusion of the  $\eta'$ field.     It is also supported by 
analysis \cite{SG} of supersymmetric models
where a similar $\theta$ vacuum structure occurs. 
The same structure also  occurs in CS phase where the  corresponding domain walls     have been  explicitly   constructed \cite{Son:2000fh}.

 \subsection{  Spontaneous symmetry breaking  of the baryon charge}\label{spontaneous}

  Second important    element  is that  in addition to this known QCD  substructures \cite{FZ, SG,Son:2000fh} of the axion domain walls expressed in terms of the  $\eta'$ and gluon    fields, there is  another substructure with a similar QCD scale which  carries  the baryon charge.   Precisely this novel feature of the  domain walls which was not explored previously  in the literature will play a key role in our proposal  because exactly  this new effect will be eventually responsible for the accretion  of the baryon charge by the nuggets. Both,     the quarks and anti-quarks   can  accrete on a given closed domain wall making eventually the quark nuggets or anti-nuggets, depending on the sign of the baryon charge.   The   sign is chosen  randomly such that equal number of quark and antiquark nuggets are formed if the external environment is CP even, which is the case when   fundamental  $\theta=0$. 
    
  Indeed, in the background of the domain wall, the physics essentially  depends on two variables, $(t, z)$. One can show  that in this circumstances  the induced baryon  charge $N$ for a single fermion  in the axion domain wall background may assume any integer value $N$, positive or negative. The total  baryon charge  $B$ accumulated on  a nugget   is determined  by  the degeneracy factor in vicinity of the domain wall    \cite{Liang:2016tqc}
    \be
   \label{N}
     N=\int d^3x \bar{\Psi}\gamma_0\Psi = -(n_1+n_2), ~~~~~~~  B=N \cdot g\cdot \int \frac{ d^2 x_{\perp}d^2k_{\perp}}{(2\pi)^2} \frac{1}{\exp(\frac{\epsilon-\mu}{T})+1}.
\ee
    In this formula      $g$ is appropriate degeneracy factor, e.g.  $g\simeq N_cN_f$ in CS phase and $\mu$ is the chemical potential in vicinity of the domain wall. 
   
   The main point of this section is that the domain walls generically  will acquire the baryon or anti-baryon charge.  This is because the domain wall tension 
    is mainly determined by the axion field while the  QCD substructure  leads to  small  correction  factor  of order $\sim \Lambda_{\rm QCD}/f_a \ll 1$.   Therefore, the presence 
    of the QCD substructure with non vanishing $N\neq 0$  increases the domain wall tension only slightly.   Consequently, this implies that the domain closed bubbles  carrying   the baryon or anti baryon charge will be copiously produced during the transition as they are very {\it  generic  configurations of  the system.} Furthermore, the baryon charge   cannot leave the system  during the time evolution as it is strongly bound to the wall due to the topological reasons. The corresponding binding energy per quark is order of $\mu$ and increases with time  as shown in \cite{Liang:2016tqc}.   One can interpret this phenomenon as a {\it local spontaneous symmetry breaking effect}, when on the scales of order  the  correlation length $\xi (T)$ the nuggets may acquire the positive or negative  baryon charge with equal probability. This is because the sign of $N$ in eq. (\ref{N}) may assume any  positive or negative values with equal probabilities. 
  
  \subsection{Kibble-Zurek mechanism}\label{KZ} 
  Next important ingredient of the proposal is the Kibble-Zurek mechanism which gives a generic picture of formation of the topological defects
 during a phase transition, see original papers \cite{KZ},  review \cite{KZ-review} and the textbook \cite{Vilenkin}.  In our context the Kibble-Zurek mechanism suggests that once the axion potential is sufficiently tilted  the  $ N_{DW}=1$ domain walls form. The potential becomes much more pronounced   when the chiral condensate forms at $T_c$. After some time after $T_a$ 
  the system is dominated by a single, percolated,  highly folded and crumpled  domain  wall network of very complicated topology. In addition, there will be a  finite portion of the closed walls (bubbles) with typical size of order correlation length $\xi(T)$, which is defined as an average  distance between folded domain walls at temperature $T$. It is known that the probability of finding closed walls with very large size $R\gg \xi$ is exponentially small.  
  
  The key point for our proposal is there existence of these finite closed bubbles made of the axion domain walls.     
   Normally it is assumed that these closed bubbles collapse as a result of the domain wall pressure, and do not play any significant  role in dynamics of the system. However, as  we already mentioned in Introduction 
 the collapse of these closed bubbles is halted due to the Fermi pressure acting inside of the bubbles.  Therefore, they may survive and serve as the dark matter candidates.
  The  percolated network  of the domain walls will decay to the axion in conventional way as discussed   in \cite{Chang:1998tb,Wantz:2009it,Hiramatsu:2012gg,Kawasaki:2014sqa}.  
 Those axions (along with the axions produced by  the conventional misalignment mechanism \cite{Wantz:2009it,misalignment})  
 will contribute to the dark matter density today.  
 The corresponding contribution to dark matter density is highly sensitive to the axion mass as $\Omega_{\rm dark}\sim m_a^{-1}$, and it is not subject of the present work.  Instead, the  focus of the present work  is  the dynamics of the 
  closed bubbles, which is normally ignored in computations of the axion production. Precisely these closed bubbles, according to this proposal,  will eventually become the stable nuggets and may serve as the dark matter candidates.  
 \exclude{ It may saturate the observed dark matter density if $m_a\simeq 10^{-6}$ eV   while it may contribute very little to  $\Omega_{\rm dark}$ if the  axion mass is slightly heavier than $m_a\simeq 10^{-6}$ eV.    
   We shall not elaborate  on the production and spectral properties of these  axions in the present work.  
 Instead, the  focus of the present paper is  the dynamics of the 
  closed bubbles, which is normally ignored in computations of the axion production.  }  
 The nugget's contribution to  $\Omega_{\rm dark}$ is not very sensitive to the axion mass, but rather,  is determined by the formation temperature $T_{\rm form}$ as explained in Introduction.

\subsection{Colour Superconductivity}\label{CS}
 There existence 
of CS phase in QCD represents the  next   crucial
element of our scenario. The   CS has been an active area of research for quite sometime,  
 see   review papers \cite{Alford:2007xm,Rajagopal:2000wf} on the subject.
 The  CS phase is realized when quarks are squeezed to the density which 
is  few times
nuclear density. It has been known that this regime may be realized in nature in neutron stars
interiors and in the violent events associated with collapse of massive stars or collisions 
of neutron stars, so it is important for astrophysics.  

 The force which squeezes quarks in neutron stars is gravity; the force which does
an analogous  job in early universe  during the QCD transition is 
a violent collapse  of a bubble of size $R\sim \xi (T)$  formed from  the axion domain wall as described  in Section \ref{KZ}  above.
 If  number density of quarks trapped inside of the bubble (in the bulk)
 is sufficiently large, the collapse 
 stops due to the internal Fermi pressure.  In this case the system starts to oscillate and the quarks 
in the bulk  may reach the equilibrium with   the ground state  being in a CS phase. As we advocate  in \cite{Liang:2016tqc}
this is very plausible fate of a relatively large size bubbles of size $R\sim \xi (T)$ made of the axion domain walls which were produced after the QCD transition. 

Indeed, one can numerically solve the equations describing the evolution of the nuggets.  A typical solution describes an oscillating bubble   with  frequency $\omega\sim m_a$. The bubble  is   slowly decreasing its radius $R(t)$ with a characteristic  dumping  scale $\tau$.   The time evolution of such a    bubble can be well approximated as follows  
 \begin{equation}
\label{eq:6.R2}
R(t)=R_{\rm form}+(R_0-R_{\rm form})e^{-t/\tau}\cos\omega t , 
\end{equation}
where $R_0$ is initial size of a bubble $R_0\sim \xi (T)$, while $R_{\rm form}$ is  a final  bible's  size when formation is almost complete. In formula  (\ref{eq:6.R2}) parameter  $\tau$ represents a typical damping time- scale which is expressed in terms of the axion mass $m_a$ and $\Lambda_{\rm QCD}$. It turns out that numerically   $\tau$ is of  order of cosmological scale $\tau\sim 10^{-4}s$. This numerical value  is fully consistent with our anticipation that the temperature of the Universe drops  approximately by a factor of $\sim 3 $ or so  during the formation period. During the same period of time the chemical potential $\mu$ inside the nugget reaches sufficiently large values  when the CS sets in.  Therefore, our phenomenological analysis  in  Section \ref{nuggets}  (where  DM nuggets are treated as    very dense objects in  CS phase) is supported   by present studies on   formation and time evolution of the nuggets made of the axion domain walls and ordinary quarks in CS phase.
  
 \subsection{Coherent $\cal{CP}$ odd axion field}\label{CP}
 If $\theta$ vanishes, then equal number of nuggets and anti-nuggets would form. 
 However,   the  $\cal{CP}$ violating $\theta$ parameter (the axion field), which is defined as value of $\theta$ at the moment of domain wall formation   generically is not zero, though it might be numerically quite small. Precisely the dynamics of the coherent axion field $\theta(t)$  leads to  preferences in formation of one   species   of nuggets, as argued in details in \cite{Liang:2016tqc}. This sign-preference   is correlated  on the scales   where the axion field $\theta(t)$ is   coherent, i.e.   on the scale of the entire Universe at the moment of the domain wall formation. 
 In other words, we assume that the PQ phase transition happened before inflation. 
 One should emphasize that this assumption on coherence of the axion field on very large scales is consistent with formation of $N_{DW}=1$ domain walls, see section \ref{DW}. 
 %This coherence   obviously cannot be satisfied for a generic type of the domains walls with $N_{DW}\neq 1$ when  $N_{DW}$ {\it  physically distinct} vacuum states with the same energy must be present in the system. 
 
Precise dynamical computations of this $\cal{CP}$ asymmetry  due to the coherent axion field  $\theta(t)$  is a hard problem of strongly coupled QCD at $\theta\neq 0$. It depends on a number specific properties of the nuggets, their evolution, their environment, modification of the hadron spectrum at $\theta\neq 0$, etc.
All these factors equally contribute to the difference between the nuggets and  anti-nuggets.  
  In order to  effectively  account for these coherent effects   one can  introduce an unknown coefficient $c(T)$ of order one
 as follows
 \be
 \label{ratio2}
 B_{\rm antinuggets}=c(T) \cdot   B_{\rm nuggets}  ,~~{\rm where  } ~~ |c(T)| \sim 1,~~
  \ee
  where $c(T)$ is obviously a negative parameter  of order one. 
% The observed ratio (\ref{ratio1}) at the end of formation corresponds to $c(T_{\rm form})\simeq -1.5$. 
 \exclude{We emphasize that the main claim  represented by eq.  (\ref{ratio2}) is not very sensitive to the axion mass $m_a(T)$ nor to the magnitude of $\theta(T)$ at the QCD transition when the bubbles start to oscillate and slowly accrete the baryon charge. The only crucial factor in our arguments is that  the  variation is correlated on the   scale where the axion field  $\theta(t)$ can be represented by the coherent superposition of the axions at rest. }
 This key relation of this framework (\ref{ratio2}) unambiguously implies that
 the baryon charge in form of the visible matter can be also expressed in terms of the same coefficient $c(T)\sim 1$
 as follows
$B_{\rm visible} =- (B_{\rm antinuggets}    + B_{\rm nuggets})$.  
Using eq. (\ref{ratio2}) the expression for the visible matter  $B_{\rm visible}$ can be rewritten as 
  \be
 \label{ratio4}
    B_{\rm visible}\equiv \left(B_{\rm baryons}+B_{\rm antibaryons}\right)
=  -\left[1+c(T)\right] B_{\rm nuggets} =-\left[1+\frac{1}{c(T)}\right] B_{\rm antinuggets}.  
  \ee
  The same relation can be also represented  in terms of the measured observables  $\Omega_{\rm visible}$ and $\Omega_{\rm dark}$ at later times  when only the baryons (and not anti-baryons) contribute to the visible component\footnote{\label{omega1}In eq. (\ref{ratio_omega}) we neglect   the differences (due to different gaps) between the energy per baryon charge in hadronic and CS phases to simplify notations.  This correction obviously does not change the main claim of this proposal stating that  $\Omega_{\rm visible}\approx\Omega_{\rm dark}$.} 
  \be
  \label{ratio_omega}
  \Omega_{\rm dark}\simeq \left(\frac{1+|c(T)|}{\left|1+c(T)\right|}\right)\cdot \Omega_{\rm visible} ~~ {\rm at} ~~ T\leq T_{\rm form}.
  \ee
One should emphasize that the relation (\ref{ratio4}) holds as long as the thermal equilibrium is maintained, which  we assume to be the case.
Another important comment   is that each individual contribution  $|B_{\rm baryons}|\sim |B_{\rm antibaryons}|$ entering  (\ref{ratio4})     is  many orders of magnitude greater  than the baryon charge hidden in the form of the nuggets and anti-nuggets at  earlier  times when $T_c>T> T_{\rm form}$. It is just their total baryon charge which is labeled as $B_{\rm visible}$ and representing the net baryon charge of the visible matter is the same order of magnitude (at all times) as the net baryon charge hidden in the form of the nuggets and anti-nuggets.

The baryons continue to annihilate each other (as well as baryon charge hidden in the nuggets) until the temperature reaches $T_{\rm form}$ when all visible anti-baryons get annihilated, while visible baryons remain in the system and represent the visible matter we observe today. It    corresponds to $c(T_{\rm form})\simeq -1.5$ as estimated below if one neglects the differences in gaps in CS and hadronic phases, see footnote \ref{omega1}. After this temperature the nuggets essentially assume their final form, and do not loose or gain much of the baryon charge from outside. The rare events of the annihilation between anti-nuggets and visible baryons continue to occur. In fact, the observational excess of radiation in different frequency  bands,    reviewed in section \ref{nuggets}, is a result of  these rare annihilation events at present time.  

The  generic consequence of this framework  represented by eqs. (\ref{ratio2}), (\ref{ratio4}), (\ref{ratio_omega})  takes  the following form   for   $c(T_{\rm form})\simeq -1.5$ which corresponds to the case when the nuggets saturate entire dark matter density today:
\be
\label{ratio5}
B_{\rm visible}\simeq \frac{1}{2} B_{\rm nuggets}\simeq -\frac{1}{3}B_{\rm antinuggets}, ~~~~~ \Omega_{\rm dark}&\simeq &5\cdot\Omega_{\rm visible},
\ee
which is identically the same relation (\ref{ratio1}) presented in Introduction. The relation  (\ref{ratio5})  emerges  due to the fact 
that  all components of matter, visible and dark,  proportional to one and the same dimensional parameter $\Lambda_{\rm QCD}$, see footnote \ref{omega1} with a comment on this approximation. In formula (\ref{ratio5})  $B_{\rm nuggets}$ and  $B_{\rm antinuggets}$ contribute to $\Omega_{\rm dark}$, while $B_{\rm visible}$ obviously contributes to $\Omega_{\rm visible}$. The coefficient $\sim 5$ in relation  $\Omega_{\rm dark}\simeq 5\cdot\Omega_{\rm visible}$ is obviously not universal, but relation (\ref{Omega}) is universal, and 
  very generic consequence of the entire framework, which   was the main motivation for the proposal \cite{Zhitnitsky:2002qa,Oaknin:2003uv}.

   \section{ Implications for the axion search experiments}\label{axion-search}
The goal of this section is to   comment on relation of our framework and the   direct axion search experiments  \cite{vanBibber:2006rb, Sikivie:2008, Asztalos:2006kz,Raffelt:2006cw,Sikivie:2009fv,Rybka:2014cya,Rosenberg:2015kxa,Graham:2015ouw}. We start with the following   comment we made in   section \ref{nuggets}:  
   this model which has a single  fundamental parameters  (a mean baryon number of a nugget $\la B\ra \sim 10^{25}$    entering  all the computations) is   consistent with all known astrophysical, cosmological, satellite  and ground based constraints as reviewed  in section \ref{nuggets}.  For discussions of this section it is convenient  to express this single  normalization parameter $\la B\ra \sim 10^{25}$  in terms of the  axion mass $m_a\sim 10^{-4}$ eV as these two parameters directly related as discussed in the origin paper \cite{Zhitnitsky:2002qa}.
   The corresponding relation between these two parameters emerges  because the axion mass $m_a$ determines the wall tension $\sigma\sim m_a^{-1}$. At the same time parameter $\sigma$  enters the expression for   equilibrium, which itself determines   the size of the nuggets, $R_{\rm form}$ (and therefore $B\sim R^3_{\rm form}$) when the  formation is complete at temperature $T_{\rm form}$. 
   
      The lower limit on the axion mass, as it is well known, is  
   determined by the requirement that  the axion contribution to the dark matter density does not  exceed the observed value $\Omega_{\rm dark}\approx 0.23$. There is a number of uncertainties in the corresponding estimates. We shall not comment on these  subtleties by referring to the   review papers \cite{vanBibber:2006rb, Sikivie:2008, Asztalos:2006kz,Raffelt:2006cw,Sikivie:2009fv,Rybka:2014cya,Rosenberg:2015kxa,Graham:2015ouw}. The corresponding uncertainties are mostly due to the  remaining discrepancies between different groups on the computations  of the  axion production  rates due to the  different mechanisms such as  misalignment mechanism versus  domain wall/string decays.  
   \exclude{In what follows to be  more concrete in our estimates we shall use   the following  expression for the dark matter density in terms of the axion mass  resulted from the misalignment mechanism \cite{Graham:2015ouw}:
 \be
 \label{dm_axion}
 \Omega_{\rm (DM ~axion)}\simeq \left(\frac{6\cdot 10^{-6} {\rm eV}}{m_a}\right)^{\frac{7}{6}}
 \ee 
 }
 If one assumes that the dominant contribution to the axion density is due to the misalignment mechanism than the estimates suggest 
 that the axion of mass $m_a\simeq 2\cdot 10^{-5}$ eV saturates the dark matter density observed today, while the axion mass in the range of $m_a\geq  10^{-4}$ eV contributes very little to the dark matter density.  
  
 There is another mechanism of the axion production when the Peccei-Quinn symmetry is broken after inflation.
 In this case the string-domain wall network produces a large number of axions such that the axion mass $m_a \simeq   10^{-4}$ eV may saturate the dark matter density,  see  relatively  recent estimates \cite{Wantz:2009it,Hiramatsu:2012gg,Kawasaki:2014sqa} with some comments and references on previous papers.    Our original remark here is that the $N_{\rm DW}=1$ domain walls 
 can be formed even if the PQ phase transition occurred before inflation and a unique physical vacuum occupies entire Universe, see section \ref{DW} with comments and references. 
   \begin{figure}[h]
\centering
\sidecaption
\includegraphics[width= 5cm,clip]{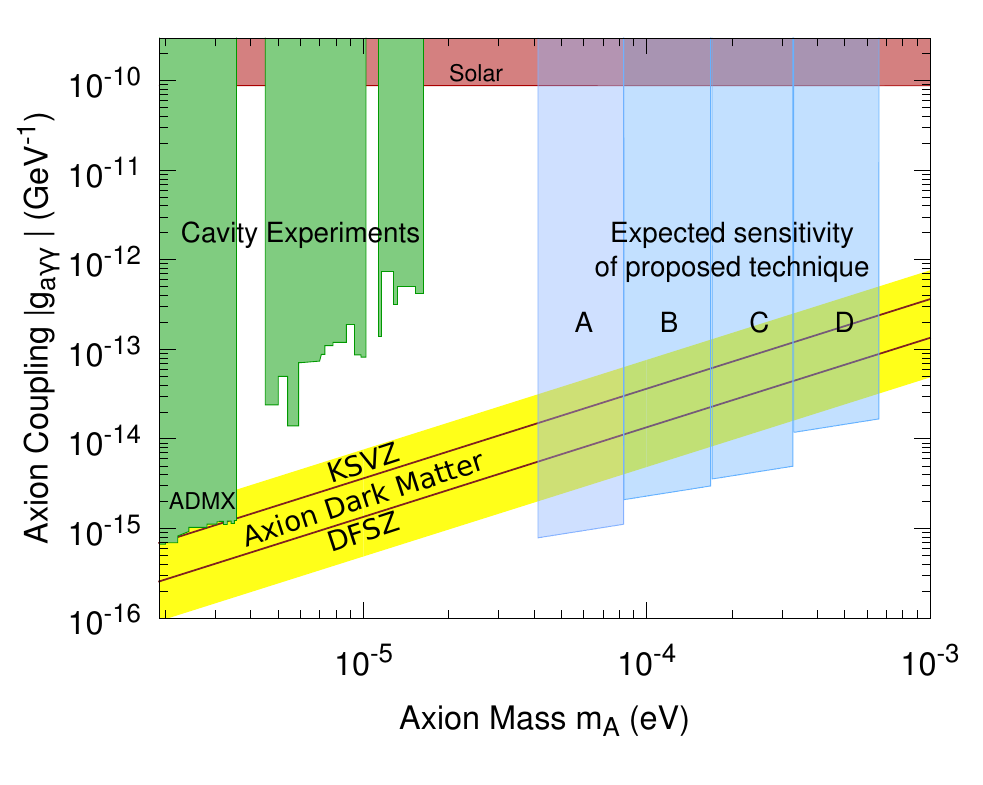}
\caption{ Cavity / ADMX experimental constraints on the axion mass shown in green.  The expected sensitivity for the  Orpheus axion search experiment \cite{Rybka:2014cya} is shown by blue regions ``A", ``B", ``C" and ``D".  In particular, experiment ``B",  covers   the most interesting region of the parametrical space with $m_a\simeq  10^{-4}$ eV  corresponding to the  nuggets with mean baryon charge $\la B\ra\simeq 10^{25}$ which itself satisfies all  known astrophysical, cosmological, satellite  and ground based constraints as discussed in section \ref{nuggets}. The plot is  taken from   \cite{Rybka:2014cya}. 
  }
\label{rybka}
\end{figure}

 The main lesson to be learnt from  the present work  is   that in addition to these well established mechanisms previously discussed in the literature there is an additional  contribution to the dark matter density also  related to the axion field. However, the mechanism which is advocated in the present work contributes to the dark matter density    through formation of the nuggets, rather than through the direct axion production.  The corresponding mechanism as argued in section \ref{CP} always satisfies the relation  $\Omega_{\rm dark} \approx \Omega_{\rm visible}$, and, in principle is capable to  saturate the dark matter density $\Omega_{\rm dark}\approx 5  \Omega_{\rm visible}$ by itself for  arbitrary magnitude of the axion mass $m_a$ as the corresponding contribution is not sensitive to the axion mass  in contrast with conventional mechanisms mentioned above.  A precise coefficient in ratio  $\Omega_{\rm dark} \approx \Omega_{\rm visible}$ is determined by a parameter  of order one, $|c(T)|\sim 1$, which unfortunately is very hard to compute from the first principles,   as discussed   in section \ref{CP}. 
 
   Our choice for $m_a\simeq  10^{-4}$ eV  which corresponds  to $\la B\ra \sim 10^{25}$ is entirely  motivated by our previous analysis of astrophysical, cosmological, satellite  and ground based constraints as reviewed  in Section \ref{nuggets}. As we mentioned  in Section \ref{nuggets}  there is a number of frequency bands where some excess of emission was observed, and this model may explain some portion, or even entire excess of the observed radiation in these frequency bands. Our  normalization $\la B\ra \sim 10^{25}$   was fixed by   eq.(\ref{flux1}) with  assumption  that the observed dark matter is saturated by the nuggets. The relaxing this assumption obviously modifies  the coefficient $c(T)$ as well as $\la B\ra$.

    Interestingly enough, this range of the axion mass $m_a\simeq  10^{-4}$ eV  is perfectly consistent with recent  claim \cite{Beck},\cite{Beck1}
  that the previously observed small signal in resonant S/N/S Josephson junction  \cite{Hoffmann} is a result of the dark matter axions with the mass $m_a\simeq 1.1\cdot 10^{-4}$ eV.  We  conclude this section on   optimistic  note with  a remark  that the most interesting region of the parametric space  with mean baryon charge $\la B\ra\simeq 10^{25}$ which corresponds to $m_a\sim10^{-4}$ eV  might be   tested by the  Orpheus axion search experiment \cite{Rybka:2014cya} as shown on Fig. \ref{rybka}.

  \section*{ Conclusion. Future directions.} 
 First, we want to list the main results of the present studies, while the comments on possible future developments will be presented at the end of this Conclusion.
 
{\bf 1.}  First key element of this proposal is the observation (\ref{N}) that the closed axion domain walls  are copiously produced and generically  will acquire the baryon or anti-baryon charge.
 This phenomenon of ``separation of the baryon charge" can be interpreted as a local version of spontaneous symmetry breaking.
 This symmetry breaking occurs not in the entire volume of the system, but on the correlation length $\xi(T)\sim m_a^{-1}$ which is determined by the  folded and crumpled axion domain wall  during the formation stage. Precisely this local charge separation eventually leads to the formation of the nuggets and anti-nuggets  serving in this framework as the dark matter component  $\Omega_{\rm dark}$. 
 
  {\bf 2.} Number density of nuggets and anti-nuggets and their size distributions will not be identically the same  as a result of the coherent (on the scale of the Universe) axion     $\cal{CP}$ -odd field.  We parameterize the corresponding effects  of order one by phenomenological parameter $c(T)\sim 1$. 
It is important to emphasize that  this parameter of order one  is not fundamental constant of the theory, but,  calculable from the first principles.  In practice, however,  such a computation could be quite a challenging problem   when even the QCD phase diagram is not  known. The fundamental consequence  of this framework, $\Omega_{\rm dark} \approx \Omega_{\rm visible}$, which is given by (\ref{Omega}) is {\it universal}, and not sensitive to any parameters as both components are proportional to $\Lambda_{\rm QCD}$.
The observed ratio (\ref{ratio1}), (\ref{ratio5}) corresponds to a specific value of $c(T_{\rm form})\simeq -1.5$ as discussed in section \ref{CP}.

{\bf 3.} Another  consequence of the proposal is a natural explanation of the ratio (\ref{eta}) in terms of the formation temperature $T_{\rm form}\simeq 40$ MeV, rather  than in terms of    specific   coupling constants which normally enter  conventional ``baryogenesis" computations. This observed  ratio is expressed in our framework in terms of a single parameter   $T_{\rm form}$ when nuggets complete their formation. 
This    parameter   is not fundamental constant of the theory, and as such is   calculable from the first principles.   In practice, however,  the  computation of $T_{\rm form}$ is  quite a challenging problem as explained in the original paper \cite{Liang:2016tqc}.
Numerically, the observed ratio 
(\ref{eta}) corresponds to $T_{\rm form}\simeq 40$ MeV which is indeed slightly below  the critical temperature $T_{CS}\simeq 60~ $MeV where the colour superconductivity sets in. 
The relation $T_{\rm form}\lesssim T_{CS}\sim \Lambda_{\rm QCD}$  is {\it universal} in   this framework as   both parameters are proportional to 
$\Lambda_{\rm QCD}$. As such, the universality of this framework  is similar to the universality $\Omega_{\rm dark} \approx \Omega_{\rm visible}$ mentioned in previous item. 
  At the same time, the ratio   (\ref{eta})  is not universal itself as it is exponentially  sensitive to precise value of $T_{\rm form}$  due to conventional  suppression factor $\sim\exp(-m_p/T)$.  

 {\bf 4.}  The only new fundamental parameter of this framework is  the axion mass $m_a$. Most of our computations (related to the cosmological observations, see section \ref{nuggets}.) however, are expressed in terms of the mean baryon number of nuggets $\la B\ra$   rather than in  terms of the axion mass. However, these two parameters are unambiguously related as explained in the text.  
 
  {\bf 5.} 
  This region of the axion mass $m_a\simeq 10^{-4}$ eV  corresponding to  average size of the nuggets $\la B\ra\simeq 10^{25}$       can be tested in   the Orpheus axion search experiment \cite{Rybka:2014cya} as shown on Fig. \ref{rybka}.

 We conclude with few thoughts on future directions within our framework. 
It is quite obvious that future progress cannot be made without a much deeper understanding of the QCD phase diagram at $\theta\neq 0$.
In other words, we  need to understand the structure of possible  phases along the third dimension parametrized by $\theta$     on Fig \ref{fig:phase_diagram}. 
 Due to the known ``sign problem",   the  conventional lattice simulations cannot be used at $\theta\neq 0$.   
Another possible development from the ``wish list"    is a deeper understanding of the closed bubble formation. Presently, very few results are available on this topic. The most relevant for our studies is the observation  made in \cite{Sikivie:2008} that a small number of  closed bubbles  are indeed observed  in numerical simulations. However, their detail properties (their fate, size distribution, etc) have not been studied yet. 
A number of related questions such as an estimation of correlation length  $\xi(T)$, the generation of the   structure inside the domain walls, the baryon charge accretion on the bubble,  etc, hopefully can be also studied in such numerical simulations. 

One more possible direction for future studies  from the ``wish list"   is a development some QCD-based  models where 
a number of hard questions   such as:  
evolution of the nuggets,  cooling rates, evaporation rates, annihilation rates, viscosity of the environment, transmission/reflection coefficients,   etc in unfriendly environment  with non-vanishing  $T, \mu, \theta$ can be addressed, and hopefully  answered.
All these and many other    effects are, in general, equally contribute  to our parameters $T_{\rm form}$ and $c(T)$  at the $\Lambda_{\rm QCD}$  scale in strongly coupled QCD. 
Precisely these numerical factors  eventually determine  the coefficients in the observed relations:  $\Omega_{\rm dark} \approx \Omega_{\rm visible}$ given by eq. (\ref{ratio_omega}) and $\eta$ given by eq. (\ref{eta}).
 
 Last but not least:  the  discovery  of the axion in the Orpheus experiment  \cite{Rybka:2014cya}  would  conclude  a long   and fascinating  journey 
  of searches for this unique and amazing particle conjectured almost 40 years ago.    Such a discovery  would be a strong motivation for related searches  of ``something else" as the axion mass $m_a\simeq 10^{-4}$ is unlikely to saturate the  dark matter density observed today.    
   We advocate the idea  that this  ``something else"  is the ``quark nuggets"  (where the axion plays the  key role in entire construction)  which could  provide the principle contribution to  dark matter of  the Universe as the relation $\Omega_{\rm dark} \approx \Omega_{\rm visible}$ in this framework  is not sensitive to the axion mass.

\section*{Acknowledgments}
 
This work was supported in part by the National Science and Engineering
Research Council of Canada.

 %
% BibTeX or Biber users please use (the style is already called in the class, ensure that the "woc.bst" style is in your local directory)
% \bibliography{name or your bibliography database}

\end{document}